\documentclass[preprint2]{aastex}

\usepackage{epsf}
\usepackage{epsfig}

\received{}
\shortauthors {Shaw et al.}
\shorttitle {The Unusual Variability of the LMC Planetary Nebula RP~916} 

\begin{document}

\title {The Unusual Variability of the Large Magellanic Cloud\\ Planetary Nebula RPJ~053059--683542
}

\author {Richard A.~Shaw\altaffilmark{1}, Armin Rest\altaffilmark{2,3}, Guillermo Damke\altaffilmark{2}, R.\ Chris Smith\altaffilmark{2}, Warren A.\ Reid\altaffilmark{4,5}, Quentin A.\ Parker\altaffilmark{4,5}}

\altaffiltext{1}{National Optical Astronomy Observatory, Tucson, AZ  85719}
\altaffiltext{2}{National Optical Astronomy Observatory/Cerro Tololo Inter-American Observatory, La Serena, Chile}
\altaffiltext{3}{Dept. of Physics, Harvard University, Cambridge, MA 02138}
\altaffiltext{4}{Dept. of Physics, Macquarie University, Sydney, NSW 2109, Australia}
\altaffiltext{5}{Anglo-Australian Observatory, PO Box 296, Epping, NSW 1710, Australia}

\email {shaw@noao.edu, arest@ctio.noao.edu, gdamke@ctio.noao.edu, csmith@ctio.noao.edu, warren@ics.mq.edu.au, qap@ics.mq.edu.au}

\begin{abstract} 

We present images and light curves of the bipolar Planetary Nebula RPJ~053059--683542 that was discovered in the Reid-Parker AAO/UKST H$\alpha$ survey of the Large Magellanic Cloud (LMC). The emission from this object appears entirely nebular, with the central star apparently obscured by a central band of absorption that bisects the nebula. The light curves, which were derived from images from the SuperMACHO project at CTIO, showed significant, spatially resolved variability over the period 2002 January through 2005 December. Remarkably, the emission from the two bright lobes of the nebula vary either independently, or similarly but with a phase lag of at least one year. 
The optical spectra show a low level of nebular excitation, and 
only modest N enrichment. Infrared photometry from the 2MASS and SAGE surveys indicates the presence of a significant quantity of dust.
The available data 
imply that the central star has a close binary companion, and that the system has undergone some kind of outburst event that caused the nebular emission to first brighten and then fade. Further monitoring, high-resolution imaging, and detailed IR polarimetry and spectroscopy would uncover the nature of this nebula and the unseen ionizing source. 

\end{abstract}


\section {Introduction} 

Variability in the nuclei of planetary nebulae (PNe) has been studied for decades, with renewed interest in the past few years in order to understand the frequency of PNe with binary central stars \citep[e.g.][]{Bond2000, MoeDeMarco06} and the role they might play in shaping the host nebula morphology and evolution. Variability within the nebulae themselves is less commonly reported, but when present has been attributed to a variety of causes. For example \citet{Feibelman_etal92} attributed changes in excitation of the nebulae IC~4997 and NGC~6572 to an increase in the temperature of the central stars (CSs); and \citet{Kato_etal01} reported two episodes of a deep fading and consequent re-brightening of the CS of NGC~2346, which they attributed to dust clouds within a PN passing in between the CS and the observer. 
More recently, \citet{Doyle_etal00} described the remarkable morphological evolution and photometric variation of M2--9, which they attributed to the interaction of ejected matter between a jet from the CS and the surrounding PN; \citet{Lutz_etal89} and later 
\citet{Corradi_etal01} argue from detailed UV and optical spectroscopy and IR photometry that He2--104 is a PN with a symbiotic (white dwarf plus a Mira variable) central star; 
and \citet{Pena_etal2004} obtained extensive UV and visual-band spectra of the complex nebula LMC--SMP~83, which they attribute to episodic bipolar ejections from a non-degenerate secondary in a close orbit with a white dwarf. 

We have initiated a study of photometric variability in Large Magellanic Cloud (LMC) PNe in order to make use of a very large, complete sample to set limits on the fraction of PN central stars that have companions of some sort, and to explore the frequency of other patterns of variability. We have used data from the extensive SuperMACHO photometric survey at CTIO of the LMC which was obtained over a ~4.2 year period from 2002 through the end of 2005 \citep[see][]{Garg_etal07}. One object, RPJ~053059--683542 (hereafter referred to by its common catalog name, RP916), was found to have spatially resolved variability, which is a highly unusual phenomenon that to date has only one other rough analog: the variability in the Galactic PN M2--9. RP916 was discovered by \citet{RP06} in the Anglo-Australian Observatory (AAO)/UKST deep H$\alpha$ survey of the LMC, who classified it as a ``true'' planetary nebula based on its morphology and spectral characteristics. It is one of the largest PNe known in the LMC, and has a bipolar morphology. Aside from the near uniqueness of the phenomenon, this type of variability is highly suggestive of an interaction of the PN central star with an otherwise undetectable binary companion. 

We describe in \S2 the extant observations for RP916, including the photometric campaign, optical spectroscopy, and broad-band IR photometry from the 2MASS catalog and from the SAGE survey with the \textit{Spitzer Space Telescope}. We analyze the characteristics of the variability in \S3, and suggest interpretations that are most consistent with the observations in \S4. 

\section {Observational Data}

Even before \citet{RP06} reported the initial discovery, a good deal of ground- and space-based data had accumulated on RP~916, including broad-band optical and infrared images, and optical spectra. Interestingly, the bulk of the observations were obtained during the course of surveys of the LMC that were designed with purposes other than PN research in mind. 


The primary optical images that were used for this study were obtained as a part of the SuperMACHO survey of a 23 deg$^2$ region centered on the bar of the LMC \citep[see][for details of the observing campaign]{Rest_etal05,Garg_etal07}. Eighty-two images in the vicinity of RP916 were obtained with the Mosaic-II CCD camera on the Blanco 4-m telescope at Cerro Tololo Inter-American Observatory between 2002 January and 2005 December. 
Most images were taken with a broad $VR$ filter with a passband that covers 510--740~nm and includes the strong nebular emission lines such as H$\alpha$ and [\ion{N}{2}] $\lambda\lambda6548, 6583$, but avoids [\ion{O}{3}] $\lambda5007$. This wide bandpass enabled the detection of faint objects ($m_{VR} \sim17.5 - 23.5$) within the typical 150~s exposures of the survey. 
The CCD plate scale is 0\farcs27 pixel$^{-1}$, which samples the PSF very well: the delivered image quality varied from roughly 0\farcs8 to 2\farcs0 over the course of the observing campaigns. An atmospheric dispersion corrector was used to minimize the differential atmospheric dispersion through the broad filter at the typical airmass of 1.3 to 1.4. 

\begin{figure}
\plotone{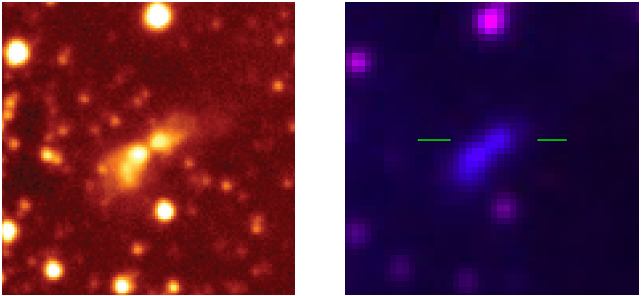}
\figcaption{\small 
Images of RP916 in the SuperMACHO $VR$ bandpass ({\it left}) with square-root intensity scale, and in the AAO/UKST H$\alpha + R$ continuum ({\it right}) with log intensity scale. Images are 30\arcsec $\times$ 30\arcsec\ with north up and east to the left. The image resolution in $VR$ is $\sim0\farcs8$, and in H$\alpha + R$ is $\sim$4\arcsec. 
The nebula is marked in the H$\alpha + R$ image with a green, horizontal bar; pure emission lines are rendered as blue, and continuum sources are rendered in pink. 
\label{fig:Images}}
\end{figure}

The SuperMACHO data were processed with an automatic pipeline \citep{Garg_etal07, Miknaitis_etal07} that corrects for electronic cross-talk between the amplifiers, removes the bias level, and applies a flat-field. 
The images were re-projected to a common geometry, with an RMS accuracy of $\approx 80$~mas. 
The PSFs are then matched, the images are corrected for sky background, and placed on a common photometric scale prior to template subtraction. 
The template image (see \S3) is shown in the left-hand side of Figure~\ref{fig:Images}, and was selected based on the good image quality ($\sim0\farcs8$) and the excellent photometric quality of the night on which this image was obtained. 

The discovery images used by \citet{RP06} were obtained as part of an AAO/UKST deep photographic survey in H$\alpha$ and in $R-$band continuum, and were obtained between 1998 and 2000. The 70~\AA\ bandwidth of the  H$\alpha$ filter also includes the [\ion{N}{2}] emission lines at $\lambda\lambda6548, 6583$. 
The photometric depth of the stacked images is $\sim21.5$ in $R$ and $R_{equiv}\sim22$ for H$\alpha$. The digitized images were sampled at 0\farcs67 pixel$^{-1}$, and the image resolution is roughly 3\farcs5. 
As described by \citet{Reid07}, the stacked H$\alpha$ and $R-$band images were each assigned a specific color in the final image: red for continuum, and blue for H$\alpha$. The discovery image of RP916 is shown in the right-hand side of Figure~\ref{fig:Images}. 

\begin{figure}
\plotone{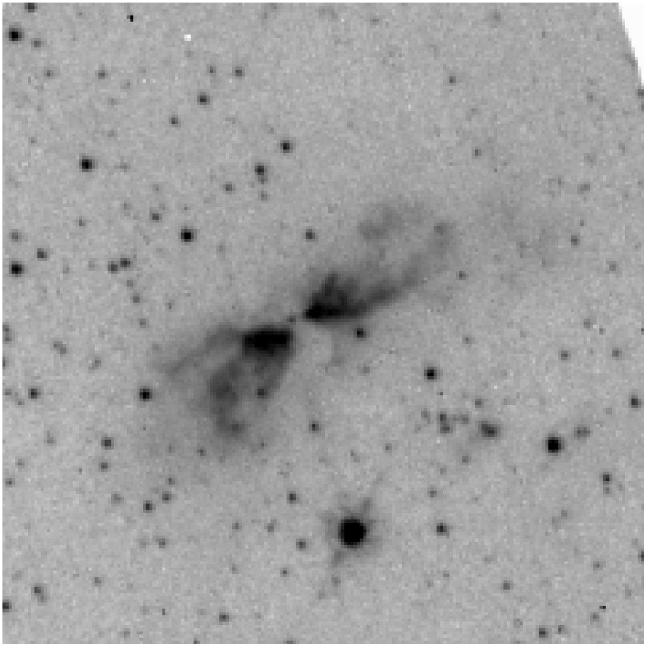}
\figcaption{\small 
\textit{HST} image of RP916 in the F814W bandpass, with log intensity scale. Image is 20\arcsec $\times$ 20\arcsec\ with north up and east to the left. 
\label{fig:HST}}
\end{figure}

Fortuitously, additional images were obtained by one of us (A.R.) with \textit{Hubble Space Telescope} using the WFPC2 camera. These images were obtained to study a microlensing target, but happened to include RP916 at the edge of the field of view. In all, four images were obtained on 2007 May 17: 
\textsc{u9px0601m} and \textsc{u9px0602m} with the F555W filter, and 
\textsc{u9px0603m} and \textsc{u9px0604m} 
with the F814W filter. The exposure times were all 500~s. We combined the calibrated images for each filter to improve the signal-to-noise ratio and to remove cosmic-rays. The F814W image is shown in Figure~\ref{fig:HST}. 


\citet{RP06} obtained two confirmatory spectra of RP916 on 2004 December 14 with the 2dF fiber spectrograph on the Anglo-Australian Telescope: one with a dispersion of 4.3~\AA~pixel$^{-1}$ 
and the other with a dispersion of 1.1~\AA~pixel$^{-1}$. 
The higher-dispersion spectrum was used to resolve 
close emission line blends
and to determine the radial velocity. 
Since the angular size of the nebula exceeds the areal coverage of the fiber, the spectra actually correspond to the western lobe of RP916. The spectra show moderately low excitation, with F([\ion{O}{3}] $\lambda5007$)/F(H$\beta$) $\sim 2.4$, relatively strong [\ion{O}{1}] $\lambda6300$ and [\ion{O}{2}] $\lambda3727$, weak [\ion{O}{3}] $\lambda4363$, no detectable \ion{He}{2} emission, and moderate [\ion{N}{2}] with F($\lambda6548$+$\lambda6583$)/F(H$\alpha$) $\sim0.5$. 


RP916 is a moderately strong infrared source (for a PN in the LMC), based on the 2MASS $J, H, K$ magnitudes \citep{Skrutskie_etal06} that were obtained 
on 2000 February 20 (MJD 51,602). It was also detected in the mid-IR SAGE survey with \textit{Spitzer Space Telescope} \citep{Meixner_etal06} in the $3.6\mu$, $5.8\mu$, and $8.0\mu$ bands with IRAC, and the $24\mu$ band with MIPS (the $70\mu$ and $160\mu$ band catalog data have yet to be published). The first epoch SAGE observations were obtained during 2005 July 15--26, around MJD 53,570. The IR brightnesses are given in Table~\ref{tab:IRphot}. 

\section {Analysis}


All of the optical images show that RP916 has a ``butterfly'' bipolar morphology \citep{Balick_Frank02}, with a dark lane (0\farcs46, or 0.11 pc, in width) bisecting the pinched main lobes of emission. The extent of the nebula is $7\farcs0 \times 3\farcs0$ ($1.72 \times 0.74$ pc at the distance of the LMC) as measured from a contour at 10\% of the peak nebular brightness in the F814W image; the extent is roughly twice as large when measured just above the sky level, with a substantial extension of faint nebular material 
to the ENE of the geometric center of the nebula. 
The H$\alpha + R$ color image shows that the emission is almost entirely nebular: what little continuum is evident in the $R-$band image undoubtedly originates from a combination of nebular continuum, emission lines in the $R$ bandpass, and the few very faint field stars that are seen in the \textit{HST} image. Interestingly, a faint star is evident in the F814W image, within the waist of RP916, offset by $\sim0\farcs23$ from the symmetry axis defined by the optical emission. The star is barely visible in the F555W image however, indicating that it is intrinsically very red, or suffers heavy extinction from dust within the waist of the nebula. It is not clear whether this is the central star of the nebula. The more detailed \textit{HST} images show clumps of emission, with the brightest knots or lobes extending almost symmetrically to the east and west but do not intersect at the geometric center of the nebula. 
The appearance of RP916 in the 2MASS images that of an extended point source, with no sign of bi-lobed structure; the object is not resolved in the \textit{Spitzer} IRAC or MIPS images. 
Although the confirmatory spectra of \citet{RP06} are neither deep nor well enough calibrated to support a detailed abundance analysis, \citet{Reid07} derived an approximate density for RP916 of $\sim400$~cm$^{-3}$ from the [\ion{S}{2}]~$\lambda6716, 6731$ line ratio. The amount of extinction, while only roughly determined, appears to be higher than average ($c\sim0.9$) for an LMC PN; given the strong IR emission, some of the extinction may be intrinsic to the nebula. Finally, the modest ratio of F([\ion{N}{2}])/F(H$\alpha$) suggests that N is not highly enriched. 
The various, general properties of RP916 are summarized in Table~\ref{tab:Properties}. 

\begin{figure}
\epsscale{1.0}
\plotone{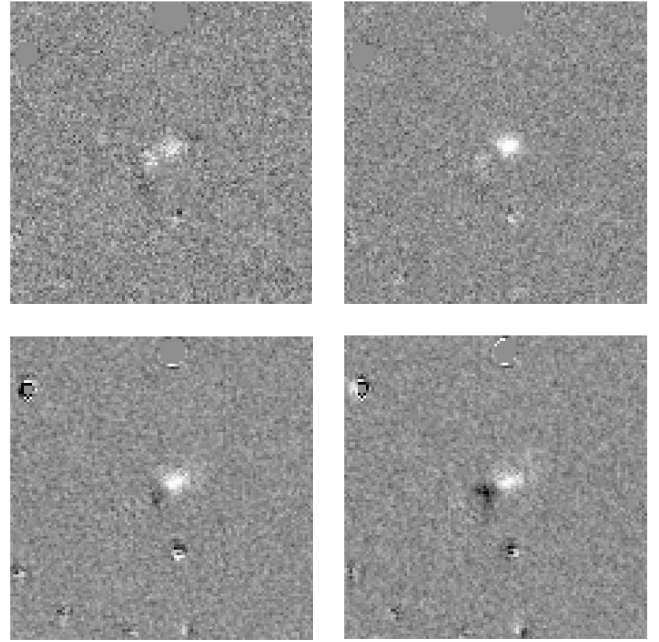}
\figcaption{\small 
Difference between the template image of RP916 taken on 2002 Jan.\ 16 = MJD~52290 (Fig.~\ref{fig:Images}) and those taken on 2002 Dec.\ 14 = MJD~52622 ({\it upper left}), 2003 Dec.\ 19 = MJD~52992 ({\it upper right}), 2004 Dec.\ 13 = MJD~53352 ({\it lower left}) and on 2005 Dec.\ 31 = MJD~53735 ({\it lower right}). Image  intensity scale is linear, with zero corresponding to medium grey. Image size and orientation as in Fig.~\ref{fig:Images}. Saturated areas in the input images are set to zero. 
\label{fig:Diff_img}}
\end{figure}

The original point of our work was to search for variability in planetary nebulae. This search was enabled by using the automated pipeline developed for the SuperMACHO project, which makes use of a very powerful difference image technique \citep[see][]{Alard_Lupton1998, Becker_etal04}. In brief, all of the images are resampled to the same geometry and sky subtracted, and the images are placed on the same photometric scale. The PSF of the template image is then matched to each target image via convolution, and subtracted. The result of this subtraction on four selected nights is shown in 
Figure~\ref{fig:Diff_img}, where the difference has the value 0.0 everywhere (apart from shot noise) except for features that are either fainter or brighter than those in the template image. Photometry is then performed at the positions of the lobes on the difference image for each observation. This technique has been shown to yield accurate photometry even in crowded fields such as the LMC, and is extremely robust against non-photometric conditions. 

Light curves were determined for the eastern and western lobes of RP916, which are shown in Figure~\ref{fig:LightCurve}. Note that both lobes increased in brightness compared to the template image, but some time after the second year of the observing campaign the eastern lobe began to fade while the western lobe continued to brighten. By MJD $\sim53,200=2004$ July the western lobe began to fade, but more slowly than the eastern lobe. Close inspection of Figure~\ref{fig:Diff_img} suggests that the orientation of the brightness enhancement in the western lobe changed and became elongated by MJD $\sim53,735=2005$ Dec. The western lobe brightened by nearly 40\% during the course of the observing campaign. 

\section {Discussion}

RP916 is unusual in many respects, including its large physical size, strong IR emission, and optical variability. But the morphology and moderate ionization spectrum leave little doubt that it is a genuine planetary nebula, and the measured radial velocity establishes its location in the LMC.
If the central star of RP916 has a binary companion that is very red and luminous it could 
contribute to the near-IR continuum luminosity, but it cannot account for the extreme colors in the mid-IR. 
Note, however, that since the epochs of the 2MASS and SAGE observations differ by 5.4 years it is possible that the ratio of the near- to far-IR luminosity could have changed somewhat during that time. 
We conclude that RP916 contains a large amount of dust (which might be expected given the pinched waist and the dark band that bisects the nebula). If RP916 is similar to other LMC PNe with significant dust, both molecular emission features and thermal continuum radiation contribute substantially to the total IR flux, particularly in the mid-IR \citep[see, e.g.,][]{Stang_etal07}. 

\begin{figure}
\epsscale{1.0}
\plotone {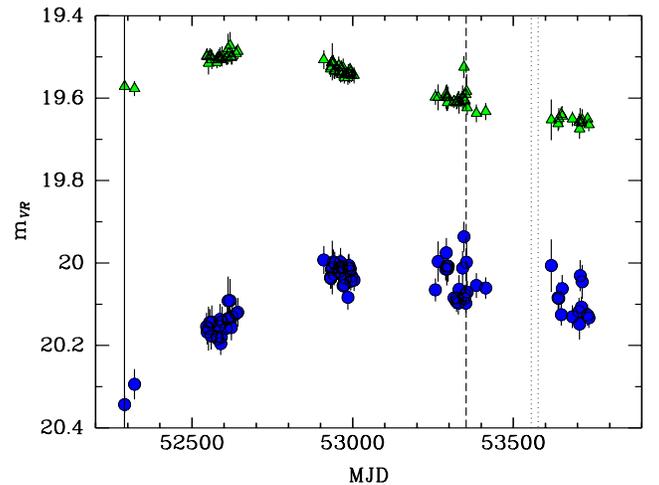}
\figcaption[f4.eps]{\small  
Light curves for eastern ({\it triangles}) and western ({\it circles}) lobes of RP916. Abscissa shows the SuperMACHO $VR$ bandpass. The epoch of the template image is indicated ({\it solid line}), as is the epoch of the optical spectrum ({\it dashed line}) and the SAGE mid-IR campaign ({\it dotted lines}). 
\label{fig:LightCurve}}
\end{figure}

The nebular variability is the most unusual feature of this object, and it is difficult to think of a cause that does not involve a binary star as the central engine. Indeed, the lack of an extreme N enrichment, which is typical of bipolar PNe in the LMC \citep{Shaw_etal2006} is consistent with common-envelope evolution, where the expected conversion of C to N via hot-bottom burning is suppressed.  It may be that the variation resulted from an outburst from a red companion. In this case a change in optical luminosity might be manifested in the nebula as a light echo seen in scattered light from dust within the nebula. (Note that the phase lag of $\sim300-500$~d between the peak emission of the two lobes in Figure~\ref{fig:LightCurve} would imply a significant inclination out of the plane of the sky.) However, no Mira-like periodicity is detected over the 1450~d observing campaign. It seems unlikely that the variation is from a global change in ionizing photons, as the inferred H recombination timescale is $\sim300$~yr if the derived density is representative of the whole nebula. In any case, a light echo cannot explain the radial structures that are evident in the detailed \textit{HST} images. The variability might on the other hand result from a precessing jet of material from the central source that interacts with (i.e., shocks) the surrounding nebula. 
Other PNe that show a similar photometric variability, resulting from a Symbiotic companion, are the Galactic bipolar PN He2--104 \citep{Corradi_etal01}, and LMC-SMP~83 \citep{Pena_etal2004}. \citet{Doyle_etal00} interpreted the variability within M2--9 as an interaction of a jet from the central source interacting with the surrounding PN, which has over time continued to shape the nebular morphology. While this mechanism is an interesting possibility for RP916, confirmation would require long-slit spectroscopy to measure the velocity field, high-resolution ($\sim0\farcs1$ or better) imaging in H$\alpha$ to reveal the time-dependent morphology of the variability, imaging polarimetry to measure the extent of any scattered light by dust, and infrared spectra to understand the nature of the IR luminosity and (possibly) detect a red, luminous companion star. 

The frequency of nebular variability generally among PNe is not known, but RP916 may be the most extreme example of how PNe morphology can be re-shaped even at very advanced stages of evolution. 
Photometric monitoring at high spatial resolution of a sizable sample of bipolar PNe in the LMC is crucial for determining what fraction of PNe experience the unusual variability of RP916. 
Perhaps the observing campaigns of the next generation of survey telescopes such as that for LSST, which will have the necessary depth, sky coverage, and cadence, can help to resolve this broader question.  

\acknowledgements 
Support for this work was provided by NOAO, which is operated by the Association of Universities for Research in Astronomy, Incorporated, under NSF cooperative agreement NAS5--26555. The SuperMACHO survey was conducted under the auspices of the NOAO Survey Program. A.R.\ thanks the Goldberg Fellowship Program for its support. SuperMACHO is supported by STScI grants GO--10583 and GO--10903. We thank the referee (K. Volk) for his extremely helpful comments. 


\clearpage

%
\begin{deluxetable}{lcclc}
\tabletypesize{\footnotesize}
\tablecolumns{5}
\tablewidth{0pt}
\tablecaption{IR Photometry \label{tab:IRphot}}
\tablehead {
\multicolumn{3}{c}{SAGE\tablenotemark{a}} & \multicolumn{2}{c}{2MASS\tablenotemark{b}} \\
\cline{1-3} \cline{4-5} \\
\colhead {Band} & \colhead {Magnitude} & \colhead {Flux (mJy)} & \colhead {Band} & \colhead {Magnitude} 
}

\startdata  
$3.6 \mu$m & $12.46\pm0.15$ & $2.9\pm0.4$ & $J$ & $16.32\pm0.18$ \\
$4.5 \mu$m & \nodata & \nodata & $H$ & $15.07\pm0.12$ \\
$5.8 \mu$m & $10.35\pm0.04$ & $8.5\pm0.3$ & $K$ & $13.88\pm0.08$ \\
$8.0 \mu$m & $8.58\pm0.03$ & $23.2\pm0.7$ & & \\
$24 \mu$m & $3.34\pm0.01$ & $340.\pm3.$ & & \\
\enddata
\tablenotetext{a}{SAGE project catalog entry J053059.45--683541.2 \citep{Meixner_etal06}}
\tablenotetext{b}{2MASS catalog entry 05305947--6835413 \citep{Skrutskie_etal06}}
\end{deluxetable} 

%
\begin{deluxetable}{lcl}
\tabletypesize{\footnotesize}
\tablecolumns{3}
\tablewidth{0pt}
\tablecaption{Basic Properties of RP916 \label{tab:Properties}}
\tablehead {
\colhead {Property} & \colhead {Value} & \colhead {Reference} 
}

\startdata  
R.A. (J2000) & $5^{\rm h}$ $30^{\rm m}$ 59\fs481 & 1 \\
Dec (J2000) & $-68\degr$ 35\arcmin\ 41\farcs29 & 1 \\
Morphology  & ``Butterfly'' bipolar & 1, 2 \\
Angular Size & $7\farcs0 \times 3\farcs0$ & 1 \\
Physical Size & $1.72 \times 0.74$ pc & 1 \\
Density &  400 cm$^{-3}$ & 3 \\
V$_{Helio}$ &  280 km~s$^{-1}$ & 2 \\
$c$ &  0.9 & 3 \\
\enddata

\tablerefs{1.\ this work; 2.\ \citet{RP06}; 3.\ \citet{Reid07} }
\end{deluxetable} 

\clearpage


\end{document}